\documentclass[%
 reprint,
nofootinbib,
 amsmath,amssymb,
 aps, amsfonts,
prd,
]{revtex4-2}
\usepackage{microtype}

\usepackage{hyperref}
\hypersetup{
    colorlinks=true,
    linkcolor=[rgb]{0,0,0.6},      
    urlcolor=[rgb]{0,0,0.6},
    citecolor=[rgb]{0,0,0.6}
    }
\setcitestyle{numbers,open={[},close={]}}
\usepackage{graphicx}
\usepackage{dcolumn}
\usepackage{bm}

\usepackage[super]{nth}
\usepackage{booktabs}
\usepackage[version=4]{mhchem}

\usepackage{Zallman,lettrine}

\newcommand{\orcid}[1]{\href{https://orcid.org/#1}{\includegraphics[width=10pt]{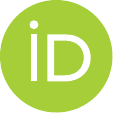}}}

\begin{document}

\preprint{APS/123-QED}

\title{Dark Matter-Powered Stars and the High-Redshift Tidal Disruption Event Rate}

\author{Thomas H. T. Wong\,\orcid{0000-0001-5570-0926}}
\email{h7wong@ucsd.edu}
\affiliation{Department of Physics, University of California San Diego, La Jolla, California 90293-0354, USA}
\author{George M. Fuller\,\orcid{0000-0002-4203-4108}}%
\email{gfuller@physics.ucsd.edu}
\affiliation{Department of Physics, University of California San Diego, La Jolla, California 90293-0354, USA}

\date{\today}

\begin{abstract}
Tidal disruption events (TDEs) result from stars being gravitationally-scattered into low angular momentum orbits around massive black holes. We show that the short lifetimes of massive Population III stars at high redshifts could significantly suppress the volumetric TDE rate because they are too short-lived to reach disruption-fated orbits.  However, this suppression can be alleviated if captured dark matter (DM) within stellar interiors provides an additional energy source, thereby extending stellar lifetimes. We find that this TDE rate revival is most pronounced for DM particles with mass $\mathcal{O}({\rm MeV})$, as this particle mass scale is optimal in the competing processes of DM accretion and evaporation in stars. 
\end{abstract}

\maketitle


In this Letter we demonstrate a surprising connection between dark matter (DM) and a frontier area of time-domain astrophysics.  The potential detection of tidal disruption events (TDEs) involving first-generation Population III (PopIII) stars by massive black holes (MBHs) offers a unique tool for probing the quiescent black hole population and the formation mechanisms for MBHs at high redshift \cite{Stone+Metzger16, KarChowdhury+24}.  We show that the early deaths of conventional massive stars could significantly suppress the TDE rate. However, stellar lifetimes could be extended in Dark Star scenarios where annihilating or decaying DM provides an energy source \cite{Spolyar+08, Iocco+08, Iocco08, Taoso+08, Natarajan+09, Freese+08evolution, Freese+09, Ripamonti+10, Stacy+14, Freese+16, Barkana18}. That would provide more time for stars to be scattered into the disruption zone, hence enhancing the TDE rate. Accumulating enough DM to effect this enhancement depends on the rest mass $m_\chi$ 
of the DM particle. Lighter particles are more readily captured in stars but easily evaporated, whereas heavier particles are not as numerous. This makes for an intriguing relationship between DM particle properties and TDE rates.

Nearly all detected TDEs are found at low redshifts, typically at $z \lesssim 1$ \cite{Komossa15, French+20, Gezari21}.  The fact that the lifetimes of these Population I (PopI) stars $(t_\star\sim10\,{\rm Gyr})$ are significantly longer than the characteristic timescale for TDE occurrence $(1/\Gamma\sim0.1-10\,{\rm Myr})$, where $\Gamma$ is the TDE rate of the host galaxy, implies that even stars initially located as far out as the MBH radius of influence, $r_{\rm inf}$, have lifetimes well exceeding the timescale for two-body scattering via orbital angular momentum exchange.  Therefore, neglecting the effect of finite stellar age on the TDE rate could be warranted. At high redshift, however, metal-poor PopIII stars with masses ranging from $30-300\,{\rm M}_\odot$ would remain on the main sequence for only $t_\star\sim2\,{\rm Myr}$, as interpolated from the standard stellar age-mass relation \cite{Schaerer02}: $\log(t_\star/{\rm yr}) \approx 3.83\,e^{-0.98\log (m_\star/{\rm M}_\odot)} + 5.91$.  The relatively short ages of these PopIII stars suggest that some of them could evolve off the main sequence while undergoing gravitational scattering toward arriving in TDE-destined orbits. These stars would not contribute to the predicted TDE rate.  Consequently, the conventional approach to deriving the standard PopI TDE rate at low redshifts likely overestimates the high-$z$ PopIII TDE rate.  

The impact of DM on stellar evolution may become significant if it accumulates in stars \cite{Freese+08evolution, Freese+16, Iocco08, Spolyar+08, Iocco+08, Natarajan+09, Taoso+08}.  Depending on the kind of DM, baryon interaction-facilitated DM capture is more likely at high $z$, as the mean ambient DM density $\rho_\chi$ scales as $(1+z)^3$ \cite{Scott+10, Raen+21}.  For example, numerous studies have proposed weakly interacting massive particles (WIMPs) as DM candidates that could be gravitationally bound in stars through tiny interactions with baryons \cite{Freese+08capture, Freese+16, Iocco08, Lopes+11, Brito+15}.  Through annihilation or decay, the DM could provide an additional power source alongside nuclear burning.  We extend the analytical results of DM capture from \cite{Neufeld+18} to account for DM accumulation in PopIII stars and to provide an estimate of how this extra fuel source could allow PopIII stars to live long enough to be scattered into the orbital parameter disruption zone.  

Here we will compute TDE rates using the loss cone (LC) mechanism \cite{Cohn+Kulsrud78, Wang+Merritt04, Stone+Metzger16, Merritt13, Merritt13book, Strubbe11, Pfister+22, Wong+22}. This mechanism describes the frequency at which a star diffuses through two-body scattering with other stars into a sufficiently low specific orbital angular momentum $L$ that it passes within the MBH tidal radius $r_{\rm t}\approx r_\star\left(M_\bullet/m_\star\right)^{1/3}$ \cite{Hills75}.  At this distance, the tidal field of the MBH with mass $M_\bullet$ overcomes the self-gravity of the star with mass $m_\star$.  The differential rate $(\#\,{\rm TDE}\,/\,{\rm yr})$ is found for each set of $(M_\bullet, m_\star, \beta)$, where $\beta=r_{\rm t}/r_{\rm p}$ is the orbital impact parameter quantifying the proximity of the pericenter distance $r_{\rm p}$ to the tidal radius $r_{\rm t}$.  

\textbf{\textit{Stellar Lifetime}} --- Our TDE rate calculation accounts for plunging orbits from all possible orbital energies $E$, but incorporates a correction term $\mathcal{P}(M_\bullet, m_\star, \beta, E)$ that accounts for the fraction of stars that still remains on the main sequence through the entire duration of the LC diffusion process. The differential TDE rate is then
\begin{equation} \label{eq:diffrate}
    \frac{d^2\Gamma}{d\ln \beta\,d\ln m_\star} = \frac{8\pi^2GM_\bullet r_{\rm t}m_\star\phi(m_\star)}{\beta}\int^{E_{\rm t}}_{E_{\rm inf}}\mathcal{G}\, \mathcal{P}\ dE
\end{equation}
where $E_{\rm t}=\left|-GM_\bullet/r_{\rm t}\right|$ and $E_{\rm inf}=\left|-GM_\bullet/r_{\rm inf}\right|$ are the magnitudes of specific orbital energies at the tidal disruption radius $r_{\rm t}$ and the radius inside of which the MBH dominates the gravitational field, the radius of influence, $r_{\rm inf}$, respectively.  In general, $r_{\rm t}\ll r_{\rm inf}$.  We employ the stellar mass function $\phi(m_\star)$ given in Ref.~\cite{KarChowdhury+24, deBennassuti+14}.  Both $\mathcal{G}$ and $\mathcal{P}$ are functions of all four variables $(M_\bullet, m_\star, \beta, E)$. $\mathcal{G}$ contains information regarding the $E$ from which most TDEs originate \cite{Cohn+Kulsrud78, Wang+Merritt04, Stone+Metzger16, Strubbe11, Pfister+22, Wong+22}. Here $\mathcal{P}$, with $0\leq\mathcal{P}\leq1$, acts as a probabilistic correction term, accounting for the likelihood of a star originating on an orbit with energy $E$ wandering into the narrow LC in orbital angular momentum space \emph{within} its lifetime $t_\star$.   For instance, Eq.\eqref{eq:diffrate} gives the standard differential TDE rate, uncorrected for ages, when $\mathcal{P}=1$. 

To modify the TDE rate for PopIII stars, we model the $L$ evolution of the target star as a random walk.  Within a single orbital period $t_{\rm orb}$, the rms change of $L$ is found to be roughly $\delta L\sim\left(t_{\rm orb}/t_{\rm relax}\right)^{1/2}L_{\rm c}$, where $t_{\rm relax}$ and $L_{\rm c}$ are the relaxation timescale and the specific angular momentum for a circular orbit with some $E$, respectively \cite{Merritt13,Merritt13book}.  Assuming an isotropic stellar velocity distribution, at any point in the orbit there exists a cone in velocity space with a half-opening angle $\theta_{\rm lc}$ pointing towards the sphere of radius $r_{\rm t}$ around the MBH (for reference, $\theta=0$ gives an exact radial plunging orbit). Any stars that diffuse into this opening angle are eliminated within a dynamical timescale \cite{Merritt13book}.  We model a random $L$ with a randomly oriented velocity vector as $\left|\Vec{L}_{\rm rand}\right|=\left|\Vec{r}\times\Vec{v}_{\rm rand}\right|=L_{\rm c}\,\sin\theta_{\rm rand}=(GM_\bullet/\sqrt{2E})\sin\theta_{\rm rand}$, with boundary conditions $\theta_{\rm lc}<\theta_{\rm rand}\leq\pi/2$ (considering only the hemisphere of all possible $\Vec{v}$) and $L_{\rm lc}< |\Vec{L}_{\rm rand}|\leq L_{\rm c}$.  We can then approximate the $L$-diffusion timescale from $L_{\rm rand}$ to $L_{\rm lc}$ as 
\begin{equation}
    \begin{aligned}
    L_{\rm rand}-L_{\rm lc} &\sim \left(\frac{t_{\rm diff}}{t_{\rm orb}}\right)^{1/2}\delta L \sim \left(\frac{t_{\rm diff}}{t_{\rm relax}}\right)^{1/2}L_{\rm c} \\
    t_{\rm diff} &\sim \left(\frac{L_{\rm rand}-L_{\rm lc}}{L_{\rm c}}\right)^2 t_{\rm relax}
\end{aligned}
\end{equation}
where $\left(L_{\rm rand}-L_{\rm lc}\right)$ is the total rms diffusion length required for a TDE and $\delta L$ is the step size in the random-walk formalism.  
\begin{figure}[h!]
    \centering
    \includegraphics[width=1.0\linewidth]{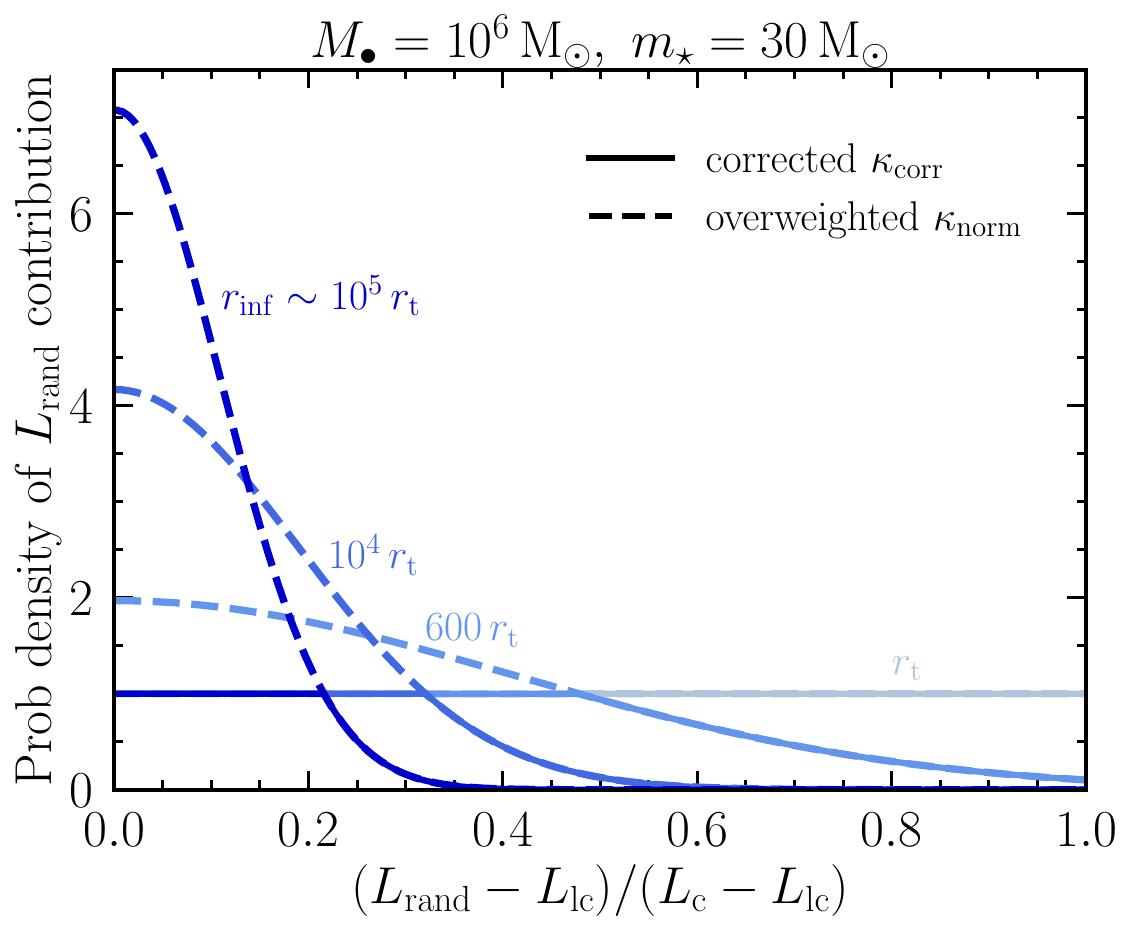}
    \caption{Probability density functions $\kappa$ of a random magnitude $|\Vec{L}|$ contributing to the TDE rate for a range of initial orbital energies $E$ for an example case with MBH mass ${10}^6\,{\rm M}_\odot$ and stellar mass $30\,{\rm M}_\odot$.  The increasing line darkness corresponds to increasing $r\propto 1/E$.  The dashed and solid curves represent the normalized $\kappa_{\rm norm}$ and the physically accurate $\kappa_{\rm corr}$ respectively.  The horizontal axis is rescaled with the maximum diffusion length.}
    \label{fig:kappadx}
\end{figure}

In this one-dimensional continuum diffusion model we can rescale both the required rms diffusion length for TDE occurrence and the corresponding number of steps, respectively, as follows: 
\begin{align}
    \frac{x}{\delta x} = \frac{L_{\rm rand}-L_{\rm lc}}{\delta L} \ \ , \ \ \frac{t}{\delta t} = \frac{t_\star}{t_{\rm orb}} \label{eq:tdt}
\end{align}
Note that all variables, except for $t_\star$, depend on the specific energy $E$.  This dependency necessitates including $\mathcal{P}$ as part of the integrand in Eq.\eqref{eq:diffrate}. To give an optimistic approximation, we allow the star to maximize its chance of scattering within $L\leq L_{\rm lc}$ by using up its entire lifetime, therefore setting $t=t_\star$ in Eq.\eqref{eq:tdt}.  The probability distribution function $\kappa$ for finding the star at $\pm x/\delta x$ (starting from the origin) after taking $t/\delta t$ steps can be taken to follow a Gaussian form: 
\begin{equation}
    \kappa\left(\frac{x}{\delta x}, \frac{t}{\delta t}\right) = \frac{1}{\sqrt{2\pi\left(t/\delta t\right)}}\exp\left[-\frac{(x/\delta x)^2}{2(t/\delta t)}\right]
\end{equation}
Since there is a finite upper limit for $L\leq L_{\rm c}$, the Gaussian distribution does not extend to infinity.  Therefore, we introduce an abrupt truncation for the maximum possible diffusion length at $x_{\rm max}/\delta x=(L_{\rm c}-L_{\rm lc})/\delta L$ and renormalize the Gaussian accordingly to get $\kappa_{\rm norm}$. 

If a massive, short-lived star is initially located far away, around $r_{\rm inf}$ from the MBH, its only potential contribution to the rate comes from the highly improbable scenario where it starts from a near-radial orbit, i.e., $L_{\rm rand}\approx L_{\rm lc}$.  At such a distant orbit, the $r_{\rm t}$ sphere resembles a ``pinhole'', with only a minuscule chance for $L_{\rm rand}$ to begin its diffusion near the edge of the very narrow LC.  In this case, although $\kappa_{\rm norm}$ is normalized to unity, the actual probability of this scenario contributing to the rate should be extremely small.  To counteract this apparent ``over-weighting'' of small $L_{\rm rand}$ values, we impose a maximum threshold for the probability density, $\kappa_{\rm corr}$.    

The overall effects are seen in Fig.\eqref{fig:kappadx} for a typical set of PopIII TDE parameters.  More $L_{\rm rand}$ from initially smaller orbits $r(E)\approx r_{\rm t}(E_{\rm t})$ can contribute to the rate (dark blue) compared to those from farther orbits $r(E)\approx r_{\rm inf}(E_{\rm inf})$ (light blue), i.e., a larger area under the curve for the $\kappa_{\rm corr}$ solid curves, as the LC (half-)opening angle $\theta_{\rm lc}$ is wider when the orbit is near the $r_{\rm t}$ sphere.  When the total required $L$-diffusion length $(L_{\rm rand}-L_{\rm lc})$ is short, the star can easily scatter into the LC.  For more distant orbits, only low $L_{\rm rand}$ can contribute significantly. The main reasons why orbits with smaller $E$ contribute insignificantly to the rate are threefold: the longer $t_{\rm orb}$ reduces the allowed number of steps within the star's limited lifetime, the longer $L$-diffusion length due to a narrower LC opening angle, and the decreased frequency of gravitational scattering from fewer field stars as stellar density drops rapidly away from the MBH.  

Finally, the total probability of a star contributing to the TDE rate is found by integrating each solid curve in Fig.\eqref{fig:kappadx}:
\begin{equation}
    \mathcal{P}(M_\bullet,m_\star,\beta,E) = \int^{x_{\rm max}/\delta x}_{0}\kappa_{\rm corr}\,d\left(\dfrac{x}{\delta x}\right)\leq1
\end{equation}
the resulting $\mathcal{P}$-factor becomes dependent solely on $E$ and no longer $L$.  See End Matter for discussions regarding the $m_\star$- and $E$-dependence of $\mathcal{P}$ for a typical $M_\bullet$ and $\beta$.  The derived $\mathcal{P}$ satisfies the following physical constraints: If $t_{\rm diff}\gg t_\star$ (indicating an initial orbit far from the MBH with $\theta_{\rm rand}\approx\pi/2$) or $t_\star\rightarrow0$ (representing a very massive star with a short lifetime), the star will most likely leave the main-sequence phase before diffusing near the LC, thus not contributing to the rate, i.e., $\mathcal{P}\rightarrow0$.  Conversely, if $t_{\rm diff}\ll t_\star$ (for an initial orbit at the edge of the tidal radius sphere or a near-radial trajectory with $\theta_{\rm rand}\approx\theta_{\rm lc}$) or $t_\star\rightarrow\infty$ (approaching the long lifetime of PopI star), the standard PopI TDE rate formalism is recovered, i.e., $\mathcal{P}\rightarrow1$.  

\textbf{\textit{DM-prolonged lifetime}} --- The culprit in PopIII TDE rate suppression is the relatively short $t_\star$ compared to the $L$-diffusion timescale $t_{\rm diff}$.  One way to alleviate the rate suppression is to consider the possibility of DM annihilation/decay processes within stars as an additional power source (DM fuel). This could extend $t_\star$ to a timescale comparable to or even longer than $t_{\rm diff}$.  This extension would allow more time for larger $L_{\rm rand}$ to diffuse into $L_{\rm lc}$.  We quantify the lifetime extension by adding an extra DM energy budget on top of the baryonic nuclear fuel, resulting in a modified stellar age expressed as 
\begin{equation}\label{eq:tageD}
    t_{\star,\chi} \sim \frac{E_{\rm baryon} + E_\chi}{L_\star} \sim t_\star + \frac{E_\chi}{L_\star}
\end{equation}
where $t_\star\sim E_{\rm baryon}/{L_\star}$ approximates the nuclear burning timescale and $E_\chi$ represents the total energy released over the star's lifetime by DM annihilation into thermalizing standard model particles.  For simplicity, we assume that the mass-luminosity relation for massive stars $L_\star\propto m_\star$ remains unchanged even after DM admixture.  

Following Ref.~\cite{Neufeld+18}, the number of DM particles retained inside a star at anytime can be expressed with 
\begin{equation}\label{eq:Nstarchi}
    N_{\star,\chi} = \frac{f_{\rm cap}\rho_\chi v_{\star,\chi}\pi r_\star^2}{m_\star}\,t\times\left(\frac{1-e^{-f_{\rm loss}t}}{f_{\rm loss}t}\right)
\end{equation}
by solving $\dot{N}_{\star,\chi} = \dot{N}_{\rm capture} + \dot{N}_{\rm loss}$. Here $\rho_\chi(E;r)$ is the ambient DM density at the initial stellar orbit, where we assume an NFW DM profile, $v_{\star,\chi}$ is the average relative speed between the surrounding DM particles and the star. We explore five DM particle masses in the range keV-GeV.  $f_{\rm cap}$ and $f_{\rm loss}$ quantify the fractions of intersected DM particles retained and lost within the star, respectively.  These fractions reflect the integrated effects of DM particles losing kinetic energy through multiple DM-baryon scattering, and the escape of DM particles stemming from atmospheric loss.  The escaping fraction due to evaporation is negligible, implying that $f_{\rm loss}t\rightarrow0$, for $m_\star\gtrsim 1\,{\rm GeV}$ as the Maxwellian velocity in the atmosphere is much slower than the stellar surface escape velocity, reducing the $t_\star$-dependence in Eq.\eqref{eq:Nstarchi} to $N_{\star,\chi}\propto t$. Our estimate aligns well with the required retained DM mass fraction necessary to significantly prolong stellar lifetimes \cite{Freese+16, John+24}.  
\begin{figure}[h!]
\includegraphics[width=0.85\linewidth]{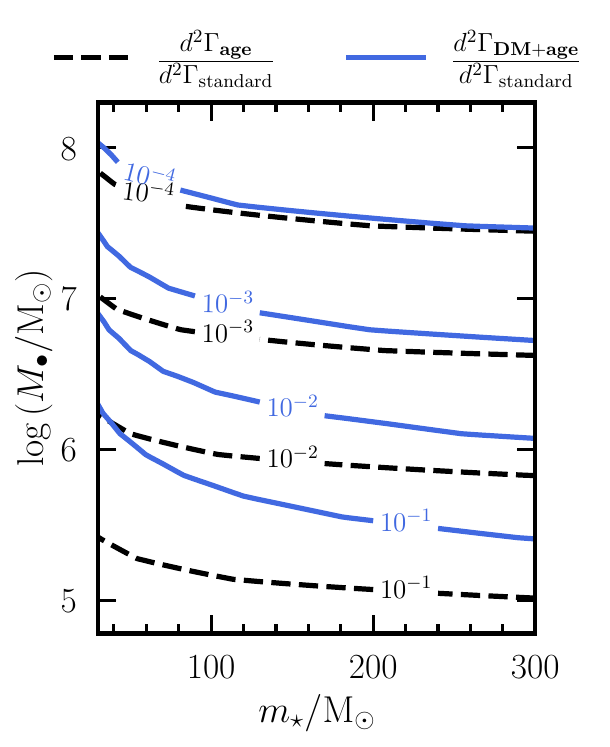}
\caption{Ratio of modified-to-standard differential TDE rates shown in the $m_\star$-$M_\bullet$ plane.  Black dashed (blue solid) contour lines show the age-to-standard (DM+age-to-standard for $m_\chi=1\,{\rm MeV}$) differential rate ratio.} \label{fig:diffrate}
\end{figure}

We average the energy released by DM annihilation at a rate per unit volume
\begin{equation}\label{eq:Qchichi}
    Q_{\chi\chi} = \frac{\langle\sigma v\rangle \rho_{\star,\chi}^2}{m_\chi} = \langle\sigma v\rangle\left(\frac{N_{\star,\chi}}{4\pi r_\star^3/3}\right)^2m_\chi c^2
\end{equation}
where we adopt $\langle\sigma v\rangle=3\times10^{-26}\,{\rm cm}^3\,{\rm s}^{-1}$ for the standard thermally-averaged WIMP annihilation cross section \cite{Freese+08evolution, Freese+16}.  Note that $\rho_{\star,\chi}$ is the mean DM density inside the star, which can and should be much larger than the mean ambient DM density $\rho_\chi$.  Integrating over the stellar lifetime and its volume yields the total energy budget contributed by DM fuel: 
\begin{equation}\label{eq:Echi}
    \begin{aligned}
    E_\chi =\ & \int^{t_\star}_0Q_{\chi\chi}\left(\frac{4\pi r_\star^3}{3}\right)dt \\
    =\ & \dfrac{\pi\langle\sigma v\rangle c^2 r_\star\left(f_{\rm cap}\rho_\chi v_{\star,\chi}\right)^2}{4m_\chi}\,t_\star^3 \times \left(\frac{1-e^{-f_{\rm loss}t_\star}}{f_{\rm loss}t_\star}\right)^2
\end{aligned}
\end{equation}
Our finding is in parallel with Ref.~\cite{Iocco+08, Taoso+08, Freese+16, John+24} where they provide a more precise estimate using the different stellar evolution codes to determine the potential increase in stellar age, they too concluded that the energy budget from DM fuel could often exceed that of the nuclear fuel.  

In the modified differential rate calculation, we replace the standard main-sequence age-mass relation $t_\star(m_\star)$ with the DM-prolonged lifetime $t_{\star,\chi}(M_\bullet,m_\star,E)$ from Eq.\eqref{eq:tageD}.  The results are presented as ratios of modified-to-standard differential rates in Fig.\eqref{fig:diffrate}. Systems with larger $M_\bullet$ and $m_\star$ in general suffer the most reduction in rate, a consequence of larger orbits with longer $t_{\rm orb}$ and shorter stellar lifetime, respectively. For example, this rate reduction can be as low as $\sim5\times10^{-5}$ for $(M_\bullet, m_\star)=(10^8\,{\rm M_\odot}, 300\,{\rm M}_\odot)$.  The black dashed contours (labeled as ``age-to-standard'', $d^2\Gamma_{\rm \mathbf{age}}/d^2\Gamma_{\rm standard}$) show the suppression in differential rate obtained by accounting for the finite ages of PopIII stars.  After incorporating DM fuel modification to the age-mass relation, the suppression changes as shown by the blue solid contours (labeled as ``DM+age-to-standard'', $d^2\Gamma_{\rm \mathbf{DM+age}}/d^2\Gamma_{\rm standard}$) for $m_\chi=1\,{\rm MeV}$.  The DM-affected suppression is milder than that of the age-corrected case.  Note that the DM-affected contours show a steeper dependence on PopIII stellar mass.  This is because the very gradual variation in the standard age-mass relation is amplified by the cubic $t_\star$-dependence in the DM energy budget term in Eq.\eqref{eq:Echi}.  
\begin{figure}[h!]
\includegraphics[width=0.95\linewidth]{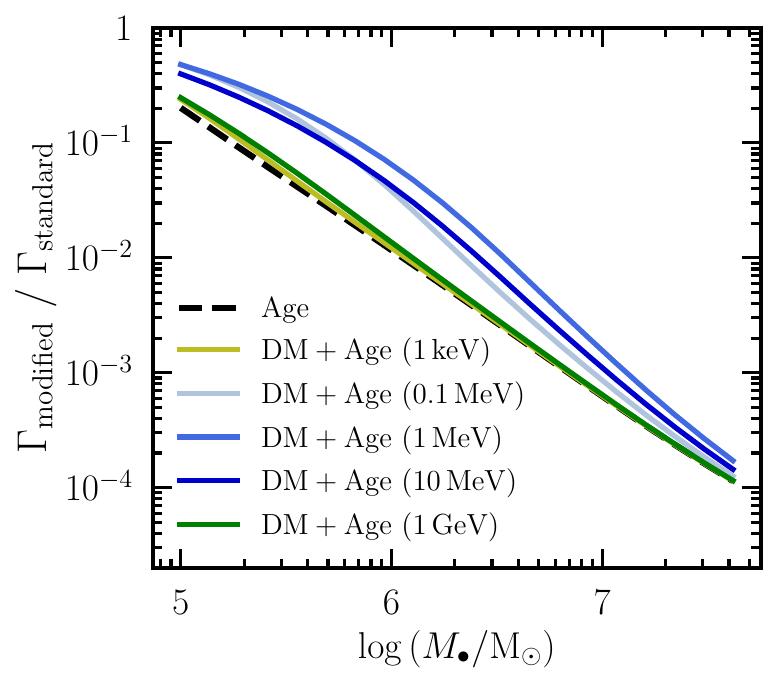}
\caption{Ratio of modified-to-standard volumetric TDE rates shown as a function black hole mass $M_\bullet$, for several cases.  The dashed curve shows the TDE rate suppression obtained by accounting only for the short PopIII stellar lifetimes. The solid curves show the milder rate suppression for PopIII stars with DM-prolonged ages, shown for different assumed DM particle masses, with $m_\chi\sim1\,{\rm MeV}$ being the least suppressed.} \label{fig:volrate}
\end{figure}

One of the key TDE observables is the volumetric rate $\Phi_\bullet\times\Gamma$ as a function of $M_\bullet$. We derive this rate by first integrating the differential rate with respect to $m_\star$ and all possible impact parameters $\beta$, and then multiplying by the black hole mass function (BHMF) $\Phi_\bullet$.  The results are shown as ratios of the modified-to-standard volumetric TDE rate in Fig.\eqref{fig:volrate}.  This ratio bypasses the uncertainties in the high-$z$ BHMF.  The rate suppression is encoded in the declining slope.  Surprisingly, the age-to-standard volumetric rate ratio closely follows a power law: 
\begin{equation} \label{eq:analyticalfit}
    \log\left(\frac{\Gamma_{\rm age}}{\Gamma_{\rm standard}}\right) = -1.24\,\log\left(\frac{M_\bullet}{10^6\,{\rm M}_\odot}\right) - 1.93
\end{equation}
As $M_\bullet$ increases, the rate suppression becomes more severe owing to the increasing contribution from stars with more distant Keplerian orbits.  The dependence on $m_\chi$ in the relative rate enhancement primarily arises from the amount of DM particles captured and retained.  For heavier DM (e.g., $m_\chi\gtrsim1\,{\rm MeV}$), the fewer number of DM particles captured stemming from low number density leads to both $E_\chi$ and $t_{\star,\chi}$ acquiring an inverse dependence on $m_\chi$ from $Q_{\chi\chi}\propto N_{\star,\chi}^2m_\chi \propto m_\chi^{-1}$.  For lighter DM, $m_\chi\lesssim1\,{\rm MeV}$, the high thermal velocities, comparable to or exceeding the escape velocity on the stellar surface, causes significant atmospheric loss causing lower DM particle retention.  Consequently, both heavier and lighter DM particles have reduced effects on prolonging stellar lifetime, making the relative rate enhancement less pronounced.  DM masses ($m_\chi\sim1\,{\rm MeV}$) between these two extremes naturally balance the deficiency of DM number density and high loss rate, resulting in optimal enhancement. Detection of DM particles with masses $\sim{\rm MeV}$ remains a frontier subject in DM direct detection \cite{Gaitskell04, Essig+12, Boddy+Kumar15, Bartels+17, Green+Rajendran17, An+18, Berlin+24}.  

In this letter we have found a significant reduction of the expected high redshift PopIII TDE rate that would stem from the short lifetimes of those stars. We have also shown that this otherwise age-suppressed TDE rate could be revived if DM annihilation powers these stars. Intriguingly, this rate enhancement depends on the rest mass of the DM particles, with $m_\chi\sim\mathcal{O}({\rm MeV})$ providing the optimal TDE rate revival. Unfortunately the DM-sensitive declining slope of the volumetric rate as a function of $M_\bullet$ introduces additional degeneracies with the highly uncertain BHMF, making it challenging to establish DM-affected TDE rates observationally.  However, next-generation all-sky deep infrared transient surveys such as the \emph{Roman Space Telescope} and NANCY \cite{NANCY+23} could reach a reduced detection rate of $\mathcal{O}(10-100)$, based on Ref.~\cite{KarChowdhury+24}.  While current \emph{JWST} deep surveys \cite{COSMOS+23, JADES+23} are unlikely to detect such events, observations of lensing cluster field could magnify the brightness of background TDEs by 1-2 magnitudes \cite{Chen+24, Szekerczes+24}, enabling shorter exposures with repeated observations to offer a more promising avenue for these transient detections. Establishing our finding of a connection between the frontier of time-domain astronomy and a fundamental issue in elementary particle physics would be profound.

\begin{acknowledgments}
We would like to acknowledge insightful discussions with S.~Arani, J.N.Y.~Chang, K.~Kehrer, S.K.~Li, and A.~Suliga. This work was supported in part by National Science Foundation (NSF) Grant  No.\ PHY-2209578 at UCSD and the {\it Network for Neutrinos, Nuclear Astrophysics, and Symmetries} (N3AS) NSF Physics Frontier Center, NSF Grant No.\ PHY-2020275, and the Heising-Simons Foundation (2017-228).
\end{acknowledgments}

\onecolumngrid
\section*{End matter}
\twocolumngrid

\begin{figure}[h!]
    \centering
    \includegraphics[width=0.95\linewidth]{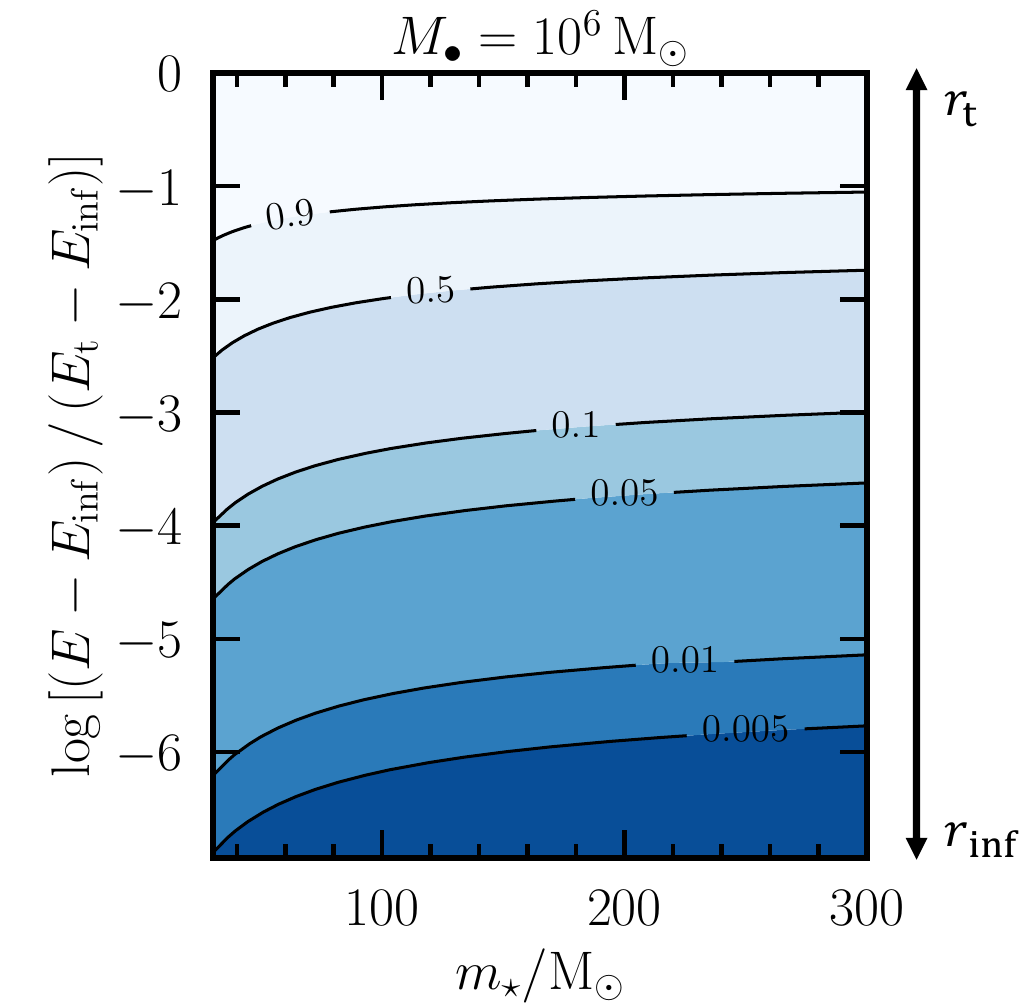}
    \caption{$\mathcal{P}$-factor as a function of $m_\star$ and $E$ for a typical MBH mass of $M_\bullet=10^6\,{\rm M}_\odot$ and $\beta=1.85$. Every value is an integration of each solid $\kappa_{\rm corr}$ curve from Fig.\eqref{fig:kappadx}. Closer orbits correspond to larger values on the vertical axis.}
    \label{fig:ProbmstarEcmap}
\end{figure}

The $m_\star$- and $E$-depedences of the $\mathcal{P}$-factor conforming to its physical constraints are shown in Fig.\eqref{fig:ProbmstarEcmap}.  This figure uses the example case of $M_\bullet=10^6\,{\rm M}_\odot$ and $\beta=1.85$.  These parameters correspond to a full disruption impact parameter.  Larger $E$, corresponding to closer orbits and shorter $L$-diffusion timescales, reproduces the standard rate estimation with $\mathcal{P}\approx1$.  Larger $m_\star$, with shorter lifetimes, have less available time to diffuse into the disruption zone. 

We reproduce the intuition that lighter stars, with longer lifetimes, tend to permit more of the larger $L_{\rm rand}$ across all ranges of $E$ since they live long enough to allow far away orbits to diffuse into the LC, e.g., stars of mass $0.3\,{\rm M}_\odot$ would yield a solid flat curve for all $r(E)$ in Fig.\eqref{fig:kappadx}, hence $\mathcal{P}$ goes to unity everywhere for all ranges of $E$.  Note that integrating $\kappa_{\rm norm}$, the dashed curves in Fig.\eqref{fig:kappadx}, would also give unity everywhere in Fig.\eqref{fig:ProbmstarEcmap}.  

For DM-fueled stars, most curves in Fig.\eqref{fig:kappadx} become flatter. The increased area under the curve of $\kappa_{\rm corr}$ in Fig.\eqref{fig:kappadx}, hence increasing the overall $\mathcal{P}$ values in Fig.\eqref{fig:ProbmstarEcmap}, indicates that the extended stellar lifetimes allow larger $L_{\rm rand}$ to contribute to the rate.  The effect of age lengthening primarily benefits the intermediate $E$ values. For these $E$ values, the ratio $\mathcal{P}_{\rm DM+age}/\mathcal{P}_{\rm age}$ peaks at $\sim 40$ for the case where $m_\chi=1\,{\rm MeV}$.  $\mathcal{P}_{\rm age}$ and $\mathcal{P}_{\rm DM+age}$ correspond to accounting for the standard finite stellar ages and the DM-fueled prolonged ages, respectively. Near $E_{\rm t}$, both $\mathcal{P}_{\rm DM+age}$ and $\mathcal{P}_{\rm age}$ plateau at unity, as stars are wandering at the edge of the LC with efficient $L$-diffusion regardless of the extended lifetimes. Near $E_{\rm inf}$, $\mathcal{P}_{\rm DM+age}\approx\mathcal{P}_{\rm age}$ because the low DM density away from the galactic center results in insufficient DM capture.

\bibliography{TDEDM}
\end{document}